\documentclass[journal=nalefd,manuscript=letter]{achemso}

\usepackage[T1]{fontenc}
\usepackage{graphicx,siunitx}
\usepackage[colorlinks=false,citecolor=blue,linkcolor=red,urlcolor=magenta]{hyperref}

\author{Valentin L.\ M\"uller}
\email{valentin.mueller@physik.uni-wuerzburg.de}
\altaffiliation{These two authors contributed equally}
\author{Yuan Yan}
\altaffiliation{These two authors contributed equally}
\affiliation{Institute for Topological Insulators and Experimentelle Physik III, Physikalisches Institut, Universit\"at W\"urzburg, Am Hubland, 97074 W\"urzburg, Germany}
\author{Oleksiy Kashuba}
\author{Bj\"orn Trauzettel}
\email{trauzettel@physik.uni-wuerzburg.de}
\affiliation{Theoretische Physik IV, Institut f\"ur Theoretische Physik und Astrophysik, Universit\"at W\"urzburg, 97074 W\"urzburg, Germany}
\author{Mohamed Abdelghany}
\author{Johannes Kleinlein}
\author{Wouter Beugeling}
\author{Hartmut Buhmann}
\author{Laurens W.\ Molenkamp}
\email{molenkamp@physik.uni-wuerzburg.de}
\affiliation{Institute for Topological Insulators and Experimentelle Physik III, Physikalisches Institut, Universit\"at W\"urzburg, Am Hubland, 97074 W\"urzburg, Germany}

\title{Electron-hole scattering limited transport of Dirac fermions in a topological insulator}

\begin{document}

\begin{abstract}
We experimentally investigate the effect of electron temperature on transport in the two-dimensional Dirac surface states of the three-dimensional topological insulator HgTe.
We find that around the minimal conductivity point, where both electrons and holes are present, heating the carriers with a DC current results in a non-monotonic differential resistance of narrow channels. 
We show that the observed initial increase in resistance can be attributed to electron-hole scattering, while the decrease follows naturally from the change in Fermi energy of the charge carriers. 
Both effects are governed dominantly by a van Hove singularity in the bulk valence band.
The results demonstrate the importance of interband electron-hole scattering in the transport properties of topological insulators.
\end{abstract}

\maketitle

A well-known prediction of Fermi liquid theory is the quadratic temperature dependence of the resistivity at low temperatures due to electron-electron scattering \cite{Kittel}.
A $T^2$ dependence of the resistivity has been experimentally observed in many alkali and transition metals, and explained by electron-electron scattering
\cite{Bass1990, Ruthruff1978, Kaveh1984}.
However, it is much more difficult to identify unambiguous fingerprints of carrier-carrier scattering in low-density semiconductor materials in which Umklapp scattering is not possible. 
Normal electron-electron scattering, on the other hand, conserves momentum and in many cases has a negligible impact on the resistivity.
As a result, the temperature-dependent resistivity in such materials is due to electron-phonon scattering \cite{Kittel}.

One of the rare counterexamples is the observation of hydrodynamic electron flow in narrow, quasi-ballistic channels made from a high-mobility 2-dimensional electron gas \cite{Molenkamp1994, deJong1995}. 
The interplay of (normal) electron-electron scattering and Fuchs-Sondheimer scattering at the partly diffusive channel walls results in a non-monotonic resistivity with increasing carrier temperature.
The initial increase of resistivity is due to the increasing rate of intercarrier collisions, which deflect otherwise parallel-moving carriers toward the dissipative sample boundaries.
This situation is similar to the Knudsen flow regime in gases \cite{Knudsen1909}.
At higher carrier temperatures, the intercarrier scattering length is sufficiently small to give rise to a laminar (Poiseuille) flow. The effective interaction with the wall is reduced and the channel resistivity decreases. 
The temperature at which the resistivity reaches the so-called Knudsen maximum is strongly dependent on the interplay between the intercarrier scattering length and the channel width.

The topological surface states of HgTe-based 3D topological insulators (3DTI) \citep{Brune2011, Brune2014, Kozlov2014} offer a new and interesting platform to investigate intercarrier scattering.
The main reason is the direct access to two-dimensional Dirac states with high mobility.
The peculiar band structure of HgTe offers the possibility to simultaneously populate topological surface states and hole-type trivial states \cite{Ziegler2020, VP2020}.
This allows to observe both n- and p-type dominated transport in the same sample by means of electrostatic gating.

In this paper, we present transport experiments in HgTe 3DTI channel structures in which we adjust the intercarrier scattering rates by changing the electron temperature.
As in our previous experiments on hydrodynamic electron flow \cite{Molenkamp1994, deJong1995}, the latter is achieved by passing a DC heating current through the sample, a technique known as ``current heating''. It is especially valuable since it allows to selectively heat the electron system while leaving the lattice temperature virtually unchanged. 
We find that the differential resistance displays a pronounced non-monotonic dependence on heating current, but only when electrons and holes are present simultaneously.
This non-monotonicity is very similar to the results in \cite{Molenkamp1994, deJong1995}, but occurs at the charge neutrality point in HgTe.
We demonstrate that the present effect is actually caused by a mechanism which is very different from the one identified in \cite{Molenkamp1994, deJong1995}.
We argue that in HgTe the initial increase in resistance can be attributed to electron-hole scattering with an appropriate cutoff temperature, while the resistance decrease can be explained by the increase of the number of highly mobile charge carriers. 
The temperature dependence of the differential resistance can well be represented taking into account these effects.
We confirm this interpretation by performing size and temperature dependent measurements.

\begin{figure*}
\centering
\includegraphics[width=.99\textwidth]{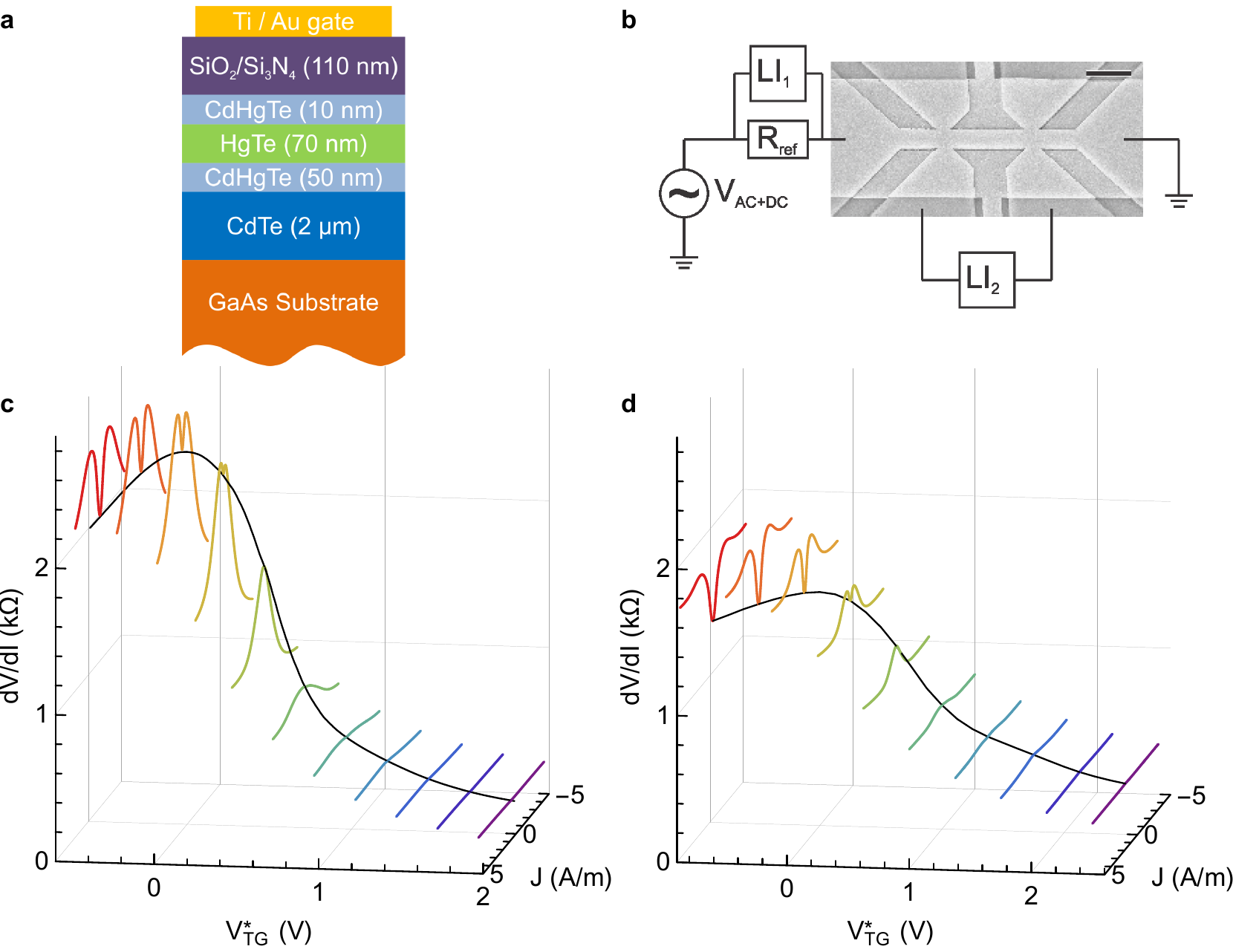}
\caption{
\textbf{a} The typical layer stack of our samples (numbers refer to sample 1), including the top gate layer.
\textbf{b} Scanning electron microscopy (SEM) image of a typical channel device, shown together with a sketch of the circuit layout, with two current contacts and four voltage probes. The top gate is recognizable by a lighter gray color. $\text{R}_\text{ref}$ is an ohmic reference resistance and $\text{LI}_1$ and $\text{LI}_2$ are lock-in amplifiers. Scale bar has a width of \SI{2}{\micro m}.
\textbf{c} and \textbf{d} Colored traces give the differential resistance dV/dI as function of heating current density $J$ at different gate voltages for sample 1 and sample 2, respectively. 
The black line represents the resistance as function of gate voltage for zero heating current at a bath temperature of \SI{1.4}{K}.
}
\label{fig:Figure1}
\end{figure*}

The experimental results are obtained on high quality 70 to \SI{75}{\nm}-thick fully strained HgTe layers grown on a relaxed CdTe layer by molecular beam epitaxy (MBE)~\cite{Brune2011,Leubner2016}.
The samples are patterned into micrometer-sized channel structures by means of a dedicated low-energy electron beam lithography process in combination with inductively coupled plasma etching. 
Subsequently, the structures are covered by a top gate stack consisting of 11 alternating \SI{10}{\nm}-thick $\rm SiO_2$ and $\rm Si_3N_4$ layers as dielectric, a \SI{5}{\nm} thick $\rm Ti$ sticking layer, and a \SI{100}{\nm} thick $\rm Au$ electrode.
The layer structure and mesa layout of a typical sample are shown in the Fig.~\ref{fig:Figure1}\textbf{a} and \textbf{b}, respectively.
The ohmic contacts to the HgTe layer are fabricated several tens of micrometers away from the channel (not visible in Fig.~\ref{fig:Figure1}).
In this paper, we present experimental results from several channel structures with different dimensions in the 1~to~\SI{100}{\micro m} range, fabricated from two wafers (cf.\ Table~S1).
Transport measurements are performed in a $\rm ^4He$~cryostat at a temperature of \SI{1.4}{\K}, if not stated otherwise.
In our experiment, the 4-terminal differential resistance of the channel is measured as a function of heating current using standard low-frequency lock-in techniques, adding a small AC voltage to the DC heating current.
Throughout the paper we refer to total carrier densities $n_\text{tot}$. The sign of $n_\text{tot}$ indicates whether the total carrier density has n-type ($n_\text{tot}>0$) or p-type character ($n_\text{tot}<0$).
Densities in n- and p-regime are obtained from Hall measurements. Densities in the transition regime are deduced by a linear interpolation.
The dependence of the electron temperature on the heating current is calibrated by Johnson-Nyquist noise thermometry for one sample.

Figures~\ref{fig:Figure1}\textbf{c} and \textbf{d} show the differential resistance dV/dI of sample 1 and sample 2 for different top gate voltages $V_\text{TG}^*$ as function of heating current density $J=I/W$, where $I$ is the DC heating current and $W$ the channel width ($W=\SI{4}{\micro m}$ for sample 1 and $W=\SI{2}{\micro m}$  for sample 2). 
The black line shows the resistance at zero current heating. It passes through a pronounced maximum when we gate the sample from n- to p-regime.
For comparison between various devices the gate voltages are presented with respect to this resistance maximum. For the two devices of Figure~\ref{fig:Figure1}, $V_\text{TG}^*=\SI{0}{V}$ corresponds to an actual applied gate voltage of $V_\text{TG}=\SI{0.8}{V}$ ($V_\text{TG}=\SI{-1}{V}$) for sample 1 (sample 2), respectively.
The total density range shown here is similar for both samples, ranging from a high n-density 
($\SI{5.7e11}{cm^{-2}}$ at $V_\text{TG}^*=\SI{2.0}{V}$ for sample 1, 
$\SI{5.3e11}{cm^{-2}}$  at $V_\text{TG}^*=\SI{2.5}{V}$ for sample 2) 
to a moderate p-density
($\SI{-2.1e11}{cm^{-2}}$ at $V_\text{TG}^*=\SI{-0.5}{V}$ for sample 1, 
$\SI{-2.0e11}{cm^{-2}}$ at $V_\text{TG}^*=\SI{-0.9}{V}$ for sample 2).
The differential resistance as function of heating current density is plotted (colored lines) perpendicular to the gate voltage axis. 
Around the resistance maximum at $V_\text{TG}^*=\SI{0}{V}$, $dV/dI$ first increases with increasing $J$, passes through a maximum and then decreases again for higher heating current densities. 
At positive gate voltages above \SI{0.5}{V}, i.\,e.\ away from the resistance maximum, $dV/dI$ decreases monotonically with heating current. 
Similar results as shown for samples 1 and 2 in Figure~\ref{fig:Figure1}\textbf{c} and \textbf{d} have been obtained for all channel samples investigated so far (cf.\,table~S1) and constitute the main experimental finding of this paper.
The peculiar non-monotonic behavior of the differential resistance is very similar to the results obtained in \cite{Molenkamp1994, deJong1995} for the crossover from ballistic to hydrodynamic electron flow. 
However, the moderate electron mobilities in our HgTe samples (cf. Table~S1) correspond to impurity scattering lengths which are shorter than the channel width.
A comparison with the numerical approach described in \cite{Kashuba2018} shows that transport of Dirac carriers should be diffusive in our case, implying that resistance changes due to intraband electron-electron scattering are small.

\begin{figure*}
\centering
\includegraphics[width=.75\textwidth]{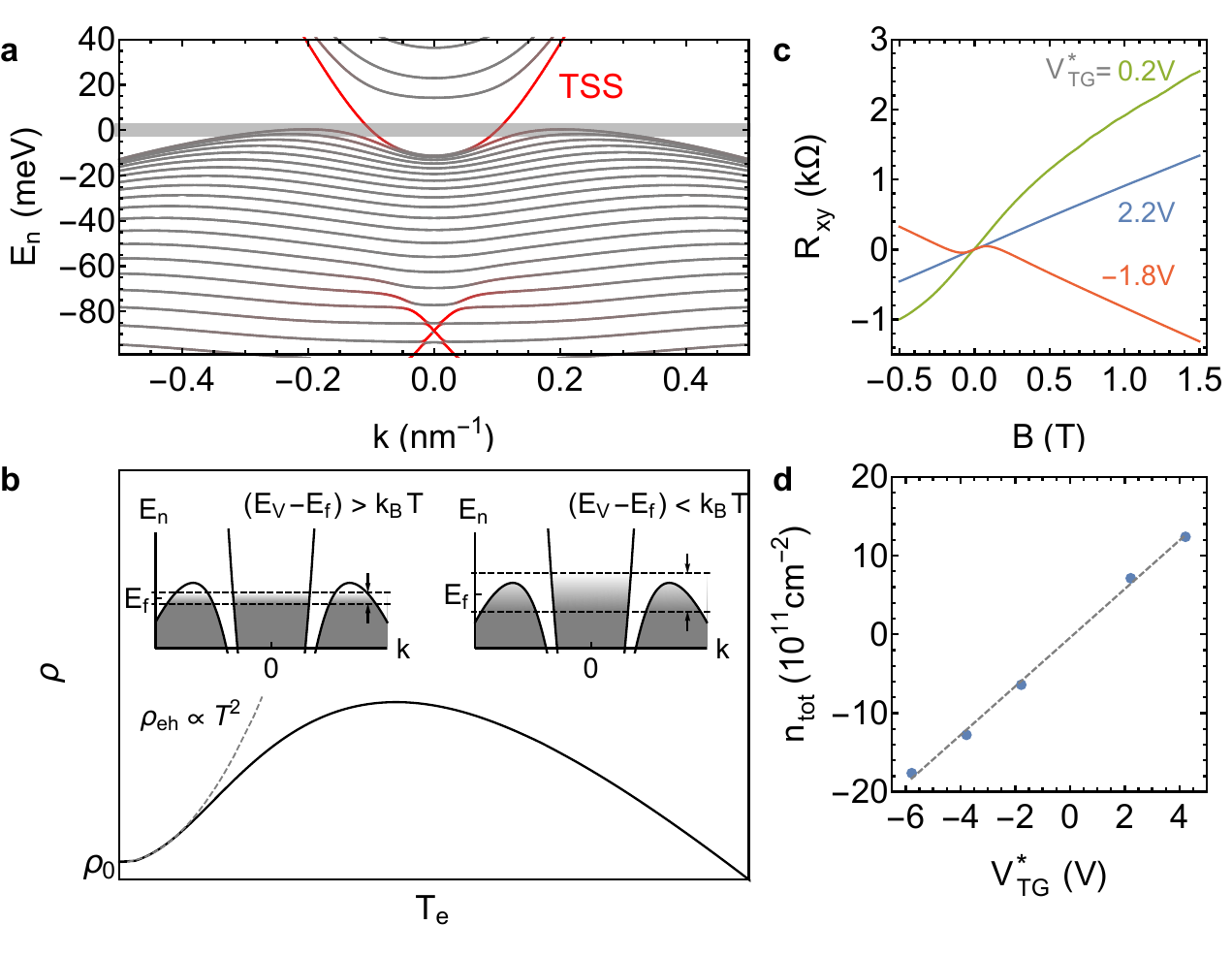}
\caption{
\textbf{a} Band structure of a \SI{70}{nm} thick HgTe layer fully strained lattice matched on a CdTe substrate, calculated in a 8-band k$\cdot$p model. The approximate energy window relevant for our model is marked by a gray shading.
\textbf{b} Schematic dependence of the resistivity $\rho$ on the electron temperature $T_e$. The inset illustrates the situation for electron-hole scattering (low $T_e$, left) and increasing electron density (high $T_e$, right).
\textbf{c} Transverse resistance $R_\text{xy}$ as function of magnetic field $B$ for three different top gate voltages $V_\text{TG}^*$ (as indicated in the figure).
\textbf{d} Total carrier density $n_\text{tot}$ as function of top gate voltage. The gray dashed line is a linear fit to the data.
}
\label{fig:Figure3}
\end{figure*}

In order to explain the observed effect, it is instructive to consider the band structure of our material system.
Figure~\ref{fig:Figure3}\textbf{a} shows the band structure of a \SI{70}{nm} thick HgTe layer grown lattice matched on a CdTe substrate in $\langle 001 \rangle$ direction, obtained by a k$\cdot$p calculation using a basis of 8~orbitals \cite{Brune2014,Novik2005}.
The figure shows the bulk bands (black). The subband splitting relates to the finite layer thickness. 
Due to the strain induced by the lattice mismatch between CdTe and HgTe, a gap of approximately \SI{14}{meV} exists between the bulk conduction band, which has $\Gamma_8^-$ character, and the valence band, which has $\Gamma_8^+$ character.
Within this gap only n-type topological surface states (TSS) are present. They disperse between the $\Gamma_8^-$ band and the $\Gamma_6$ band, which is located deep below the Fermi level (not visible in Figure~\ref{fig:Figure3}\textbf{a}).
Consequently, the Dirac point is buried \SI{90}{meV} deep into the valence band. Note the flat maximum at finite $k$ in the valence band, often referred to as ``camel back''.
This feature results from the inverted band ordering of the $\Gamma_6$ and $\Gamma_8^-$ bands and hybridization with the $\Gamma_8^+$ band which is located between them \cite{Brune2014, Shamim2020a, VP2020}. 
Such hybridization-induced van Hove singularities commonly occur in topological materials (see e.\,g.\ \cite{Yan2017}).

From the presence of the camel back with a high density of states (DOS), one expects that the Fermi level $E_f$ will be pinned to the valence band edge when gating the system from n- to p-regime. 
Once the Fermi level is pinned, electrons and holes are present in the system simultaneously. 
This behavior is experimentally confirmed by the Hall measurements shown in Figure~\ref{fig:Figure3}\textbf{c} for sample 1.
At large positive gate voltages the Hall resistance is linear (cf.\ the \SI{2.2}{V} trace in Figure~\ref{fig:Figure3}\textbf{c}).
The total carrier density $n_\text{tot}$ at this gate voltage is of order $\SI{10e11}{cm^{-2}}$ and electron-like, cf.~Figure~\ref{fig:Figure3}\textbf{d}.
At small positive gate voltages (cf.\ the \SI{0.2}{V} trace in Figure~\ref{fig:Figure3}\textbf{c}), we observe a deviation from linear behavior in the Hall voltage, indicating the onset of two-carrier behavior.
Note that this is the same gate voltage region where the dip appears in the dV/dI traces of Fig.~\ref{fig:Figure1}.
At strongly negative gate voltages, where the total carrier density is in the high p-regime, the Hall resistance still shows a pronounced two-carrier behavior (cf.\ the \SI{-1.8}{V} trace in Figure~\ref{fig:Figure3}\textbf{c}), consistent with the expected pinning of the Fermi level at the camel back.
The relevant energy range for the strongly non-monotonic dV/dI is thus the region around the valence band maximum (cf. gray shading in Figure~\ref{fig:Figure3}\textbf{a}), where we have both electrons and holes in our channel.

In such a two-carrier system, increasing the carrier temperature by current heating has two effects.
1) When free electrons in topological surface states and free holes in the bulk coexist, electron-hole scattering increases the resistivity since the low mobility bulk holes act as momentum sinks \cite{Entin2013, Nam2017}. 
Since scattering is restricted to the energy interval of $k_BT$ around the Fermi level, the intercarrier scattering time in a Fermi liquid is inversely proportional to the carrier temperature, $\tau_{eh}^{-1} \propto T^{2}$.
This mechanism aligns with our experimental observation that the differential resistivity increases with increasing current (i.\,e.\ carrier temperature).
We tentatively assign this observation with momentum relaxation due to electron-hole scattering.
2) At higher current (carrier temperature) the differential resistivity decreases. To understand this behavior, one needs to consider the phase space for electron-hole scattering.
The insets of Fig.~\ref{fig:Figure3}\textbf{b} illustrate the thermal smearing for low (left) and high (right) temperature. 
The phase space for electron-hole scattering cannot increase infinitely.
Once the thermal smearing (of order $k_BT$) reaches the valence band edge, the available phase space is truncated and the electron-hole scattering rate saturates. 
At the same time, the hole concentration continues to increase with carrier temperature when the Fermi level is sufficiently close to the valence band edge ($|E_V-E_f|\leq k_B T$). Also the number of highly mobile electrons increases so as to keep the total charge (which is determined by the gate voltage) constant.
These combined effects lead to a decrease in resistivity and the $dV/dI$ trace develops a characteristic maximum. 
The constant carrier density obviously also implies a shift of the Fermi level as a function of carrier temperature \cite{Kittel}.
This effect is strongest for gate voltages where the Fermi energy is in the bulk band gap and in close proximity to the valence band edge. In this situation electron-hole scattering is absent and the differential resistance only decreases with increases heating current (cf.\ Figure~\ref{fig:Figure1}). 
When the Fermi energy is further removed from the valence band edge (by gating to large positive gate voltages), the decrease in differential resistance becomes less pronounced.

We now will use the band structure calculation of Fig.~\ref{fig:Figure3}\textbf{a} to quantify the above qualitative considerations.
We first calculate both electron, $n_e(T)$, and hole, $n_h(T)$, concentrations for the total charge densities $n_{tot}$ found in the experiment and electron temperatures $T$ in the range of \SIrange{0.4}{30}{K}.
The intrinsic mobility for electrons $\mu_e$ is calculated from the experimental low-temperature resistivity at zero heating current, where for simplicity we assume the holes to be localized \cite{Shamim2020a}.
The temperature-dependent resistivity can then be calculated as
\begin{equation}
\rho_{tot} (T) = \frac{1}{e n_e \mu_e} + \rho_{eh}(T),
\end{equation}
where $\rho_{eh}$ is the excess resistance due to electron-hole scattering. 
It can be calculated using the coupled kinetic equation approach described in \cite{Maldague1979} as
\begin{equation}
\rho_{eh}(T) = \frac{m_e m_h n_h}{e^2 n_e (m_e n_e + m_h n_h) \tau_{eh}(T)},
\end{equation}
where $m_e$ and $m_h$ are the effective masses of electrons and holes, respectively. 
As mentioned above, the electron-hole scattering rate $\tau_{eh}(T)$ is expected to follow a $T^2$-dependence at low temperature and saturate at higher temperatures.
Numerical studies, which deliberately avoid the usual assumptions leading to the $T^2$ dependence, show that the typical temperature scale for saturation is given by $(E_V-E_f)/k_B$ \cite{Giamarchi1992, Swift2017}.
We hence apply a heuristic expression for the electron-hole scattering rate:
\begin{equation}
\tau_{eh}^{-1}(T) = C T^2 \left[ 1 - \exp\left(-\frac{E_V-E_f}{k_B T}\right) \right]^2, \quad \text{for} \quad E_f<E_V
\label{eq:ehscatteringrate}
\end{equation}
The proportionality constant $C=\SI{4e9}{s^{-1}K^{-2}}$ is estimated by comparison to our experimental results and kept fixed for all curves.
For $E_f>E_V$, where experimentally no dip is observable in the differential resistivity, we neglect the contribution of electron-hole scattering entirely.
The conversion from $\rho(T)$ to $dV/dI (I) $ then requires a translation of electron temperature into heating current. For this, we use a noise calibration done by means of Johnson noise thermometry on a sample which is comparable to sample 1 (for details, see Supporting Information).

\begin{figure*}
\centering
\includegraphics[width=.74\textwidth]{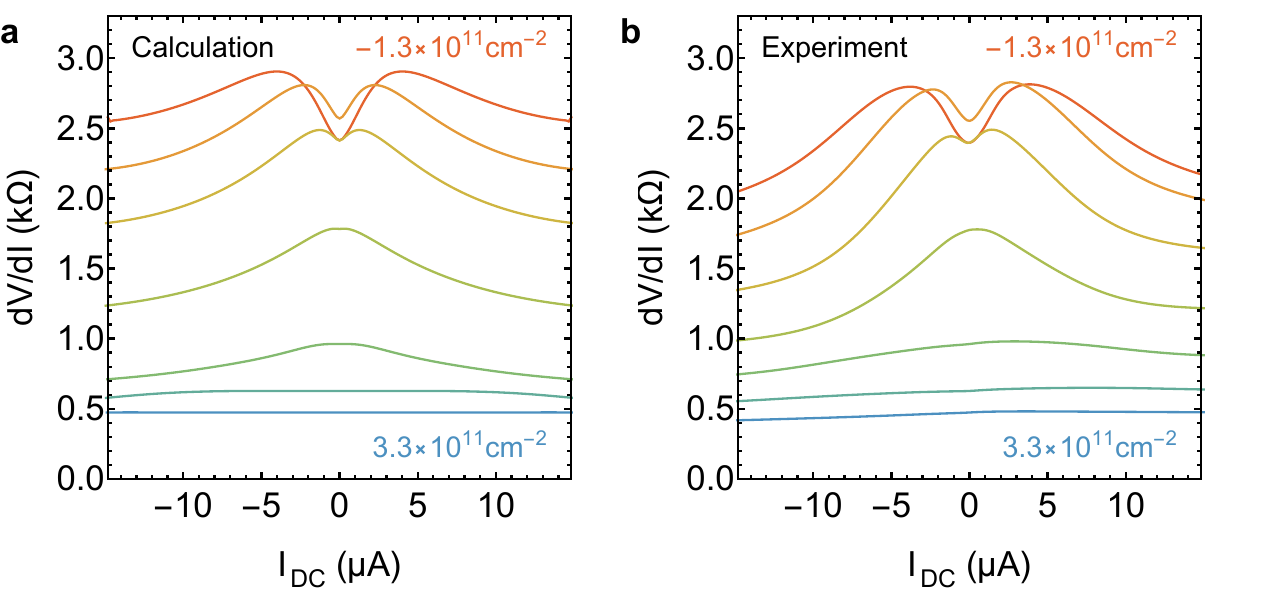}
\caption{
Comparison between heuristic model and experimental results of sample 1.
a) Calculated $dV/dI$ as function of heating current $I$ for different total carrier densities.
b) Experimental $dV/dI$ traces as function of heating current $I$. (Same data as shown in Figure~\ref{fig:Figure1}\textbf{c}).
Both figures show the same set of carrier densities.
}
\label{fig:Figure4}
\end{figure*}

Figure~\ref{fig:Figure4}\textbf{a} shows the numerical result in comparison with the experimental data for sample 1 (Fig.~\ref{fig:Figure4}\textbf{b}).
The model reproduces all aspects of the experimental data quite well.
Our model quantitatively explains the magnitude of the resistance changes and thus indicates that the observed $dV/dI(J)$ behavior is determined by heating of the charge carriers and the resulting carrier-carrier scattering.
We find that the order of magnitude of the electron-hole scattering rate constant $C$ (Eq.~\ref{eq:ehscatteringrate}) agrees well with values found in a previous study on electron-hole scattering in a semimetallic HgTe quantum well \cite{Entin2013}.

We have performed several further experiments to validate our interpretation of the observed non-monotonic behavior of $dV/dI(J)$.

\begin{figure*}
\centering
\includegraphics[width=.75\textwidth]{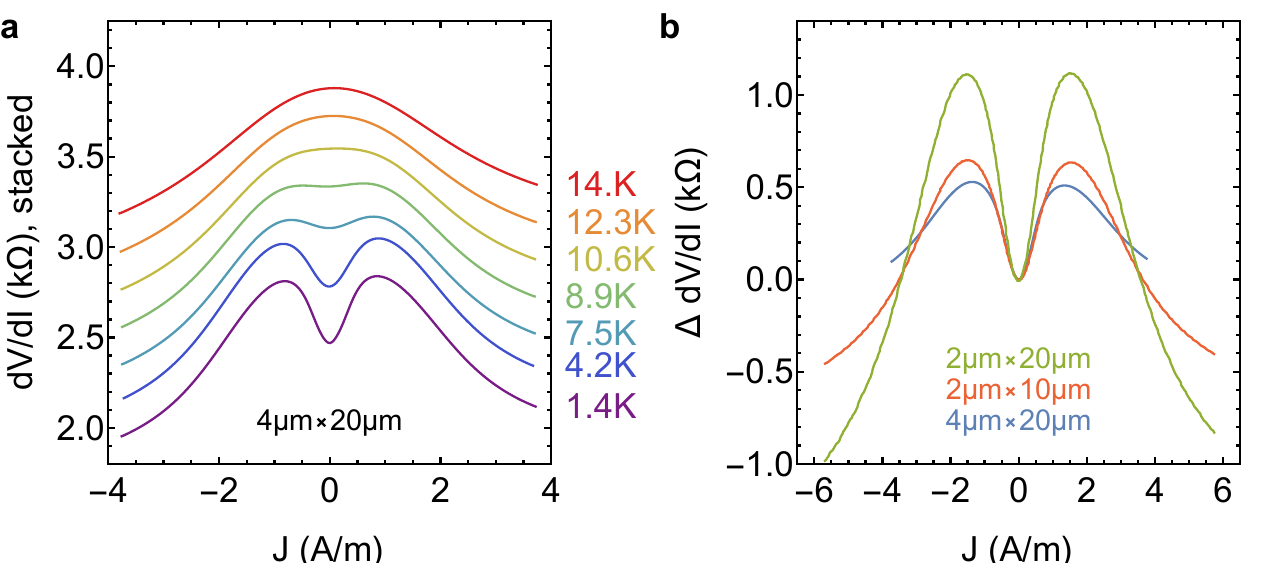}
\caption{
\textbf{a} (Bath-) temperature dependence of dV/dI for sample 1.
An offset of $\SI{0.2}{k\Omega}$ is added with increasing temperature for clarity.
\textbf{b} Differential resistance change $\Delta dV/dI = dV/dI - dV/dI (J=0)$, as function of heating current density $J$ for differently sized channel structures  as indicated in the figure.
The \SI{2}{\micro m} wide channels were measured at a bath temperature of \SI{2.5}{K}.
}
\label{fig:Figure2}
\end{figure*}

First, we demonstrate that the non-monotonicity indeed is a carrier temperature-induced effect and is not directly caused by the increased bias.
Figure~\ref{fig:Figure2}\textbf{a} shows the differential resistance of sample 1 at a fixed gate voltage of $V_\text{TG}^*=\SI{-0.2}{V}$ for several sample temperatures in the range \SIrange{1.4}{14}{K}.
With increasing sample temperature the zero-bias resistance increases, so that a peak in the differential resistance is only observable up to $T=\SI{8.9}{K}$. For higher temperatures, i.\,e.\ $T=\SI{10.6}{K}$ and above, we do not observe a non-monotonic behavior, but rather an overall decrease in differential resistance with current.
This observation confirms our reasoning that the resulting carrier temperature is responsible for our observations.
Furthermore, we can infer a lower bound of $T=\SI{8.9}{K}$ for the electron temperature at which the resistance peak occurs, which is consistent with our modeling.

Second, we verify that the observed changes of the differential resistance only depend on current density, as appropriate for the mechanism discussed above.
Figure~\ref{fig:Figure2}\textbf{b} compares the change in differential resistance $\Delta dV/dI = dV/dI(J) - dV/dI (J=0)$ as function of heating current density $J$ for sample 1 and sample 3, which differ in widths $W$ and lengths $L$ (cf.\ Table~S1). All measurements are done at the same total carrier density of  $\SI{-2.1e11}{cm^{-2}}$, where the differential resistance displays a pronounced peak at a finite current density. 
Sample 3 has additional voltage probes to do a 4-terminal measurement inside the channel (length $L=\SI{10}{\micro m}$) or across the entire channel ($L=\SI{20}{\micro m}$), possibly including entrance effects. 
The results show that the amplitude of the differential resistance changes is roughly proportional to the aspect ratio $L/W$ (i.\,e.\ the number of squares) of the channel samples.
Furthermore, comparison of the channels with $W=\SI{4}{\micro m}$ and $W=\SI{2}{\micro m}$ shows that the peak position occurs at approximately the same current density and thus shows hardly any effect of the channel width. 
The same holds for samples 6 through 8 (Supporting Information, Figure~S1).
Notably, the absence of a strong width dependence provides further proof that the non-monotonic temperature dependence of the differential resistance in our HgTe samples is not caused by the transition from the ballistic to the hydrodynamic flow regime as in high-mobility (Al, Ga)As channel samples \cite{Molenkamp1994, deJong1995}.
There, the position and amplitude of the Knudsen maximum are strongly dependent on the interplay between the (intraband) electron-electron scattering length and the channel width, which determine the effective interaction with the wall.
This is clearly not the case for the non-monotonic differential resistance observed in our HgTe samples.

Finally, in \cite{Raichev2017} it is suggested that the Bloch-Grüneisen electron-phonon scattering regime can also lead to a non-monotonic temperature dependence of the resistivity. 	
However, this can be excluded for our experiments, where all electron-phonon relaxation occurs after the hot electrons leave the channel.
In addition, assuming a typical electron density of $\SI{1e11}{cm^{-2}}$ per surface state and a velocity of sound of \SI{2.6}{km/s}\cite{Sreckovic2003}, we find a Bloch-Grüneisen temperature of only \SI{4.5}{K}, which is too small to fit our results. Furthermore, the resistance increase should become more pronounced and shift to higher electron temperatures as we increase the electron density -- quite the opposite of what we observe.

To summarize, we have investigated the differential resistance of HgTe 3DTI layers by means of current heating and observed a pronounced non-monotonic behavior when the Fermi level is tuned close to, or into, the bulk valence band.
We find that interband scattering between electrons in the topological surface states and bulk holes causes an increase in resistance, while a carrier-temperature induced increase of mobile electrons at higher currents leads to a decrease in resistance.
A model based on the band structure of our HgTe layers shows that both effects are determined by the van Hove-singularity in the bulk valence band.
Our results underline once more the importance of the bulk bands in understanding transport properties of topological surface states. 

The direct observation of electron-hole scattering in thick HgTe layers is remarkable, since it adds a new material class (i.\,e.\ 3D topological insulators) to a small list of low-carrier density systems, in which intercarrier scattering has been proven to considerably affect transport properties.

\begin{acknowledgement}

We thank C.\,Ames and L.\,Lunczer for growing the sample material.
We acknowledge financial support from the Deutsche Forschungsgemeinschaft (DFG, German Research Foundation) in the Leibniz Program and in the projects SFB~1170 (Project ID 258499086) and SPP~1666 (Project ID 220179758), from the EU ERCAdG program (Project 4-TOPS), from the Würzburg-Dresden Center of Excellence `Complexity and Topology in Quantum Matter' (CT.QMAT), and from the Free State of Bavaria (Elitenetzwerk Bayern IDK `Topologische Isolatoren' and the Institute for Topological Insulators).

\end{acknowledgement}

\begin{suppinfo}

Sample overview, further size-dependent data, and detailed explanation of our Johnson noise thermometry setup.

\end{suppinfo}

\section{Author contributions}

H.B. and L.W.M. planned the project and the design of the experiments. 
V.L.M. and Y.Y. conducted the transport experiments and analyzed the data.
V.L.M. fabricated the devices. J.K. oversaw the device fabrication.
Y.Y. and M.A. built the Johnson-Nyquist noise thermometry setup. Y.Y. performed the noise measurements and data analysis.
V.L.M., Y.Y., O.K., B.T., H.B., and L.W.M. developed the model to interpret the data.
V.L.M. and O.K. performed the simulations. 
W.B. provided the code for the band structure calculations.
B.T., H.B., and L.W.M. supervised the project.
All authors contributed to paper writing.

\bibliography{EH_sc_Paper}

\end{document}


\onecolumngrid
\newpage

\setcounter{equation}{0}
\setcounter{figure}{0}
\setcounter{section}{0}

\renewcommand\thesection{\Roman{section}}
\renewcommand\theequation{S\arabic{equation}}
\renewcommand\thefigure{S\arabic{figure}}
\renewcommand\thetable{S\arabic{table}}

\begin{center}
{{\bf\large Supporting Information

 ``Electron-hole scattering limited transport of Dirac fermions in a topological insulator''}}
\end{center}

\section{Sample overview}

Table~\ref{tab:sampleoverview} provides an overview of all samples presented in this work.

\begin{table*}[b!]
\centering
\begin{tabular}{c | c | c | c | c }
Name		&	Wafer	&	\multicolumn{2}{c|}{Wafer characterization at $V_\text{TG}*=\SI{1}{V}$}			&		Channel dimensions	\\
 			&			&	$n_\text{tot}$  ($\SI{}{10^{15} cm^{-2}}$)	&		$\mu_\text{av}$ ($\SI{}{10^{3} cm^2/Vs}$)	&			(width $\times$ length)							\\\hline
Sample 1	&	Q2823	&  2.0	&		170		&	$\SI{4}{\micro m}\times\SI{20}{\micro m}$			\\
Sample 2	&	QC0509	&  2.4	&		330		&		$\SI{2}{\micro m}\times\SI{10}{\micro m}$ 			\\
Sample 3	&	Q2823	&  2.0	&		170		&		$\SI{2}{\micro m}\times(\SI{10}{\micro m}\text{  or  }\SI{20}{\micro m})$				\\
Sample 4	&	Q2823	&  2.0	&		170		&		$\SI{4}{\micro m}\times\SI{20}{\micro m}$ 			\\
Sample 5	&	QC0509	&  2.4	&		330		&		$\SI{2}{\micro m}\times\SI{20}{\micro m}$ 			\\
Sample 6	&	QC0509	&  2.4	&		330		&		$\SI{1}{\micro m}\times\SI{10}{\micro m}$ 			\\
Sample 7	&	QC0509	&  2.4	&		330		&		$\SI{0.5}{\micro m}\times\SI{5}{\micro m}$ 			\\
\end{tabular}
\caption{
Overview of channel samples used for this work, refering to the wafer number and the channel dimensions as well as to the characterization data obtained from large Hall bar structures ($\SI{200}{\micro m}\times\SI{600}{\micro m}$ in size).
}
\label{tab:sampleoverview}
\end{table*}

\clearpage

\section{Width-dependence in samples 5, 6, and 7}

\begin{figure*}[b!]
\centering
\includegraphics[width=.49\textwidth]{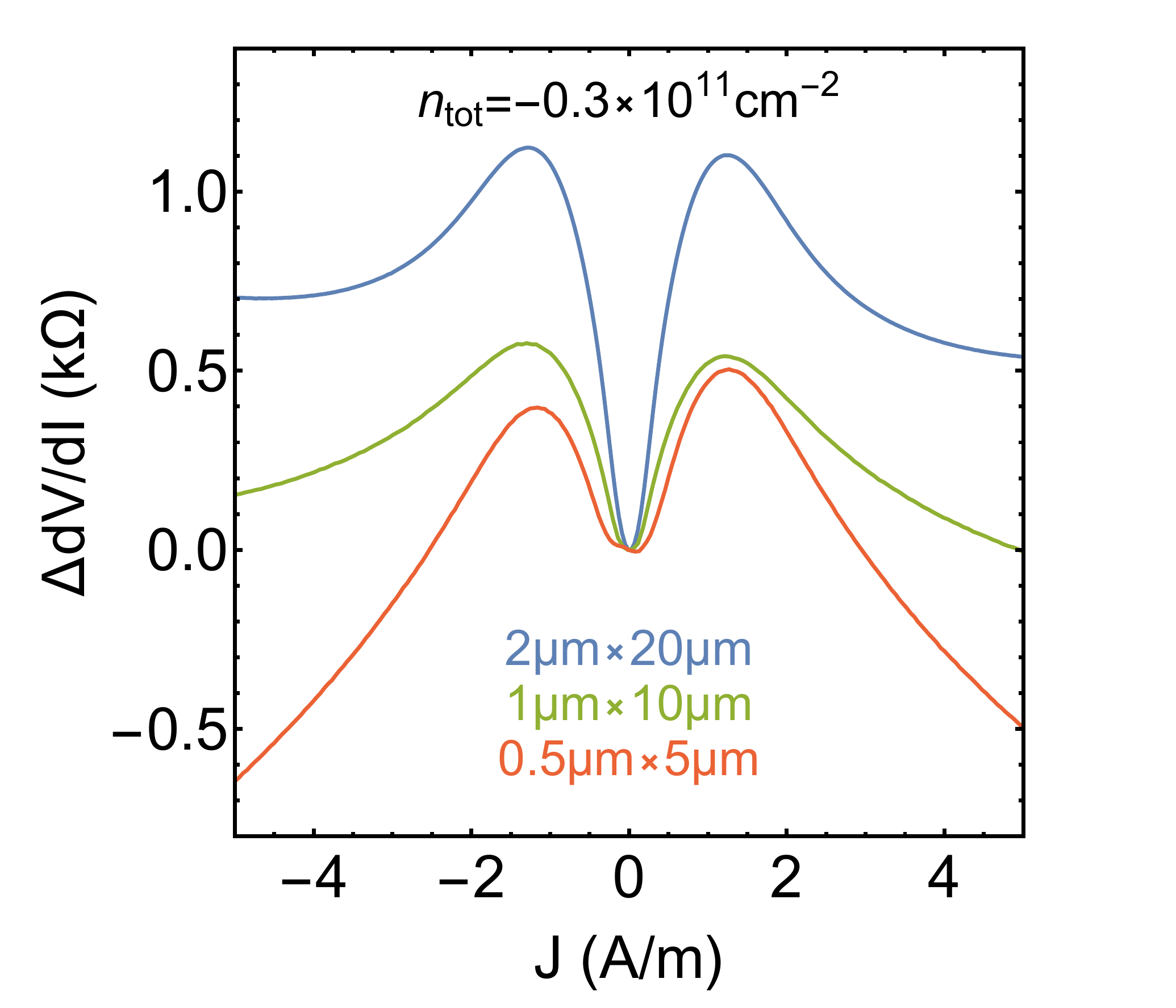}
\caption{
Width scaling in samples 5, 6, and 7. Differential resistance change $\Delta dV/dI = dV/dI - dV/dI (J=0)$ as function of heating current density for three different channel samples with different widths, but same aspect ratio (dimensions indicated in the figure). 
}
\label{fig:FigureS2}
\end{figure*}

Here we present additional measurements for size scaling.
Figure~\ref{fig:FigureS2} shows the current-heating induced changes in differential resistance for three different channel structures fabricated from wafer QC0509 at a total carrier density of $\SI{-0.3e11}{cm^{-2}}$. Similar to the experiments discussed in Figure~4b, the peak in $\Delta dV/dI$ appears at approximately the same current density, even though the channel width ranges from \SI{2}{\micro m} to \SI{0.5}{\micro m}. 

\clearpage

\section{Electron temperature calibration}

In order to find the temperature range relevant for our experiments we need to know the relation between heating current and electron temperature. This is important for translating calculated $\rho$ vs. $T$ curves to $dV/dI$ vs. $I$ curves. 
By applying a DC bias current ($I_{\rm DC}$) through the channel with resistance $R$, a Joule heating power $P=I_{\rm DC}^{2}R$ is injected into the electronic system.
At low sample temperatures a temperature difference between the electron system and the phonons (or bath) will be introduced, since the electron-phonon interaction is weak.
The electron temperature is monitored by Johnson noise thermometry \cite{DiCarlo2006, Crossno2015}.
As for a quasi-ballistic system, the applied current is expected to create a constant effective electron temperature within the channel.
For high bias voltages ($eV \gg k_B T$), the current noise can be approximated by the relationship 
$S_I = 4 k_B T_{\rm e}/R$, 
where $T_{\rm e}$ is the effective electron temperature and $k_{B}$ is the Boltzmann constant \cite{Nagaev1995}. 
We use this relationship to find an estimate of $T_{\rm e}$.

\begin{figure}
\centering
\includegraphics[width=0.9\textwidth]{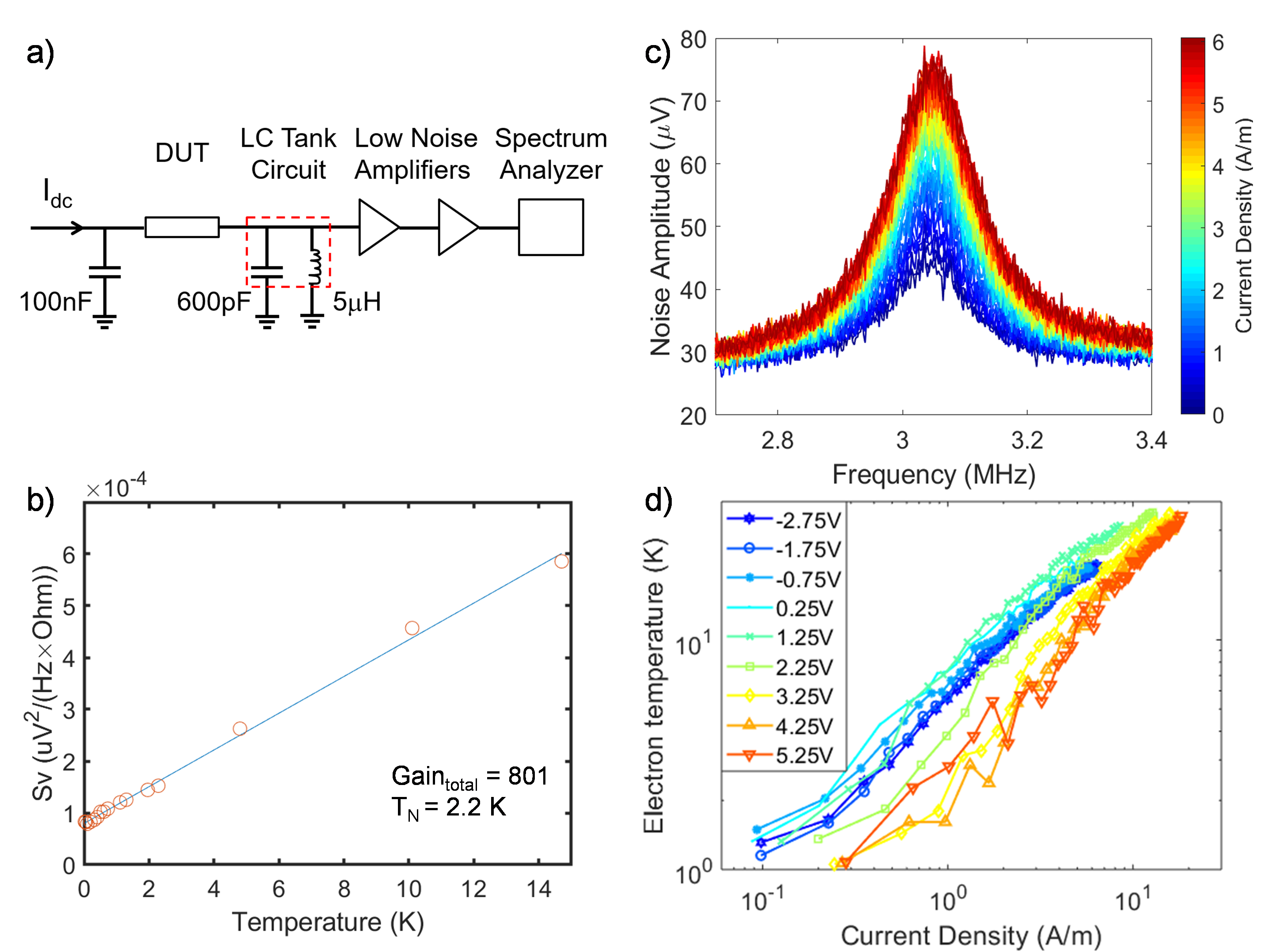}
\caption{%
a) Schematics of Johnson noise thermometry setup used to measure the electron temperature in the device under test (DUT).
b) The power spectrum density of a thin metal film (Ti/Au \SIlist[list-units = single, list-pair-separator = /]{5;80}{nm}) resistor plotted against bath temperature. The solid line is a linear fit to the data.
c) The noise spectrum of sample 4 under different DC bias currents at a top gate voltage of $V_\text{TG}^*=\SI{-2.75}{V}$.
d) The representative electron temperature and DC bias current relationship of sample 4 at different top gate voltages (given w.\,r.\,t.\ the resistance maximum).
The bath temperature is \SI{1.1}{\K}. 
}
\label{fig:sm:TInoise}
\end{figure}

The experimental setup is shown in Fig.~\ref{fig:sm:TInoise}a. 
One end of the device under test is connected to ground by a \SI{100}{\nano\farad}-capacitor to create a high frequency short.
The other end of the device under test is connected to an $LC$ tank circuit to select the required frequency band. This is followed by a home built low noise amplifier which is kept at the 1-K pot temperature to reduce its intrinsic noise. The signal is further amplified by another low noise amplifier and measured by a spectrum analyzer.
The Johnson noise thermometry setup is calibrated with the help of a metal film resistor at different bath temperatures, which allow us to extract the gain ($G$) of the system and its equivalent noise ($T_{\rm N}$) (see Fig.~\ref{fig:sm:TInoise}b). 
The relation between the measured noise amplitude $S_{\nu}$ and the electron temperatur $T_{\rm e}$ is evaluated using the following equation: $S_{\nu}=G^{2}\times 4k_{B}(T_{\rm e}+T_{\rm N})R$.

We applied Johnson noise thermometry to measure the $T_{\rm e}$ for sample~4, an exact copy of sample~1.
The measured thermal noise and converted electron temperatures at different DC bias currents and gate voltages are shown in Figure~\ref{fig:sm:TInoise}c and d. 
These calibration curves are used to convert the calculated $\rho(T)$ curves to $\rho(I)$. 
It is then straightforward to obtain $dV/dI(I)= R + I dR/dI$ with $R(I)=(L/W) \rho(I)$.

The consistency of the Johnson noise thermometry was tested in relation with temperature dependent measurements of SdH-oscillations.

\bibliography{EH_sc_Paper}